\documentclass{elsarticle}
\usepackage[version=3]{mhchem}
\usepackage{mathcomp}
\usepackage{float}
\usepackage[english]{babel}
\makeatletter
\begin{document}
\begin{frontmatter}
\title{Limits on the source properties of FR-I galaxies from high-energy neutrino and gamma observations}
\author{Isaac Saba }
\address{Ruhr-Universität-Bochum, 44780 Bochum, Germany}
\author{Julia Becker Tjus}
\address{Ruhr-Universität-Bochum, 44780 Bochum, Germany}
\author{Francis Halzen}
\address{Department of Physics, University of Wisconsin, Madison, WI-53706, USA}
\begin{abstract}
Active galactic nuclei (AGN) are believed to be the source of ultra high energy cosmic rays (UHECRs, $E>10^{18}\,\textrm{eV}$). Particles are assumed to be accelerated in the accretion disk and the plasma jets, produced due to conservation of angular momentum, to the highest energies, where they interact with each other and produce pions, which decay among others in neutrinos.
\newline
For a known cosmic ray spectral behavior, the main parameters in the calculation of the neutrino flux from proton-proton interactions are the target density $n_{\textrm{H}}$ and the ratio of electrons to protons $f_{\textrm{e}}$. Using most recent neutrino flux limits from IceCube point source searches, we set limits on the target densities  for 33 FR-I galaxies. The densities are shown to be smaller than 30 cm$^{-3}$ to $2\cdot 10^{3}$ cm$^{-3}$, depending on the source and when using a fixed electron to proton ratio of $f_e=0.1$. This implies that some cosmic ray acceleration sites, especially those close to the core of the AGN, can already be excluded, or else that the ratio of electrons to protons deviates significantly from the commonly used value of 0.1.
\newline
For Centaurus A (Cen~A) and Messier 87 (M 87) we use Fermi observations to model the $\gamma$-flux, the neutrino flux and the resulting target density.
The detection of these neutrinos will help to find information about acceleration processes in the source.
\end{abstract}
\begin{keyword}
Active galactic nuclei, FR-I galaxies, Inelastic proton-proton interaction, Target density, Centaurus A, Messier 87
\end{keyword}
\end{frontmatter}
\section{Introduction}
Active Galactic Nuclei (AGN) are the most powerful permanent objects known. The observed luminosities  range from $10^{41}$ erg s$^{-1}$, measured for nearby galaxies,  up to $10^{47}$ erg s$^{-1}$ for distant galaxies. The prevalent picture is a supermassive black hole (SMBH), located in the center of the host galaxy, with gravitational energy as the source of the luminosity.
\newline
The emission is spread widely across the electromagnetic spectrum, often peaking in the ultra-violet, but with significant luminosity in the X-ray and infrared bands. The emitted power varies on time scales of years, days or minutes. Due to angular momentum conservation, plasma is ejected and forms bipolar collimated  jets,  strong radio sources if the host galaxy is elliptical, or weak radio sources if the host galaxy is a gas rich spiral.
AGN are already observed and identified as sources of high energy $\gamma$-rays and additional observations indicate that the arrival direction of the highest energetic CR  might correlate with the position of Cen A and M 87 \cite{cena,reimer}.
\newline
High energy neutrinos in AGN are produced in coincidence with high energy $\gamma$-rays when high energetic protons interact with a local target and produce secondaries, which decay among others into neutrinos. The exact mechanism of the energy and momentum transfer in AGN is still under debate. Matter which is attracted by the SMBH, the central engine of the AGN cannot directly fall into the black hole, since it possesses angular momentum. A disk  of matter is formed within which magnetic viscosity transfers angular momentum outward and mass inward. In the core region protons can be accelerated to high energies via reconnection of the magnetic field to high energies \cite{Bednarek}. The highly energetic protons can interact  with other protons and photons, producing secondary particles as high energetic neutrinos, which can leave their point of origin almost unimpeded to be detected on Earth. A detailed approach to proton-photon and proton-proton interaction in the jet can be found in \cite{becker}.
\newline
Here we focus on  proton-proton (p-p) interaction in the core region and concentrate on the muon neutrino flux calculation for 33 FR-I galaxies. Assuming that the electrons lose all their energy to synchrotron radiation, the radio luminosity is equal to the electron luminosity
\begin{align}
L_{e}\approx L_{\textrm{radio}}.
\end{align}
Furthermore  protons and electrons are expected to be accelerated   at the same site, meaning that the proton luminosity can be determined by assuming a constant ratio $f_{e}=L_{\textrm{e}}/L_{\textrm{p}}$ between radio and proton luminosity \cite{krause}. Using this assumption, the proton luminosity can be estimated from radio observations of individual sources. A further astrophysical parameter is the target density $n_{\textrm{H}}$, which determines the optical depth $\tau_{\textrm{pp}}$ of the p-p interaction. It is given by
\begin{align}\label{e function}
\frac{I}{I_{0}}=\exp(-\tau_{\textrm{pp}}),
\end{align}
where $I_{0}$ is the initial intensity and $I$ the observed one. For the calculation the optical depth is given by
\begin{align}
\tau_{\textrm{pp}}=R\cdot n_{\textrm{H}}\cdot \sigma_{\textrm{inel}}(E_{\textrm{p}}),
\end{align}
where $n_{\textrm{H}}$ is the target density, $\sigma_{\textrm{inel}}(E_{\textrm{p}})$ the inelastic cross section for  proton proton interaction and $R$ the is the size of the interaction region. For the considered parameter space (emission from the core region of the AGN, $R<100\,\textrm{kpc}$, with densities $n_{\textrm{H}}<2\cdot10^{3}$~cm$^{-3}$), Equation (\ref{e function}) can be approximated by a linear behavior of the optical depth.
\newline
For Cen A and M 87 there are detailed $\gamma$-ray observations available, which will be used to normalize the $\gamma$-flux and to determine the neutrino flux and the resulting target densities $n_{\textrm{H}}$ at a given fraction $f_{\textrm{e}}$ \cite{abdo}.
\newline
This paper is constructed in the following manner. In section 2 the main calculation will be introduced.
In section \ref{FR1}, the results for the neutrino fluxes and the derived limits on the target density for 33 FR-I galaxies will be presented. The observations of $\gamma$-rays from Cen A and M 87 provides the possibility of deriving an exact value for $n_{\textrm{H}}$, which is done in section \ref{fermI}. Conclusions are presented in section \ref{conclusion}.
\section{Modeling neutrino and $\gamma$-ray spectra from FR-I using the sources radio luminosity}
High energy neutrinos can be produced through inelastic proton-proton interactions, where high energy protons interact with ambient protons and produce pions which decay into neutrinos. The muon  neutrino flux at Earth is given by \cite{kelner}
\begin{align}\label{flux}
\phi^{\nu_{\mu}}(E_{\nu})=\epsilon_{\textrm{osc}} \cdot c\ n_{\textrm{H}}\int_{0}^{1}{\sigma_{\textrm{inel}}\frac{\textrm{d} N_{\textrm{p}}}{\textrm{d}A\,\textrm{d}E_{\textrm{p}}}(E_{\nu}/x)F_{\nu}(x,E_{\nu}/x)}\frac{\textrm{d}x}{x} \ \textrm{with} \ x=E_{\nu}/E_{\textrm{p}}.
\end{align}
Here, $\epsilon_{\textrm{osc}}$ considers the oscillation, $n_{\textrm{H}}$ is the target density in $\textrm{cm}^{-3}$, $\sigma_{\textrm{inel}}$ is the inelastic proton-proton  cross section in mb, given by
\begin{align} \label{sigma}
\sigma_{\textrm{inel}}=(34.3+81.88L+0.25L^{2})\left[1-\left(\frac{E_{\textrm{th}}}{E_{\textrm{p}}}\right)^{4}\right]^{2} \ \textrm{with} \ \ L=\ln\left(\frac{E_{\textrm{p}}}{1 \ \textrm{TeV}}\right),
\end{align}
with $E_{\textrm{th}}$ the threshold energy for the $\pi^{+}$ production.
Neutrinos at the sources are created with the ratio
\begin{align}
\left((\nu_{\textrm{e}}+\overline{\nu}_{\textrm{e}})+(\nu_{\mu}+\overline{\nu}_{\mu})+(\nu_{\tau}+\overline{\nu}_{\tau})\right)=(1:2:0)
\end{align}
Due to oscillation of neutrinos from the distant source to Earth the ratio to be expected here is
\begin{align}
\left((\nu_{\textrm{e}}+\overline{\nu}_{\textrm{e}})+(\nu_{\mu}+\overline{\nu}_{\mu})+(\nu_{\tau}+\overline{\nu}_{\tau})\right)=(1:1:1),
\end{align}
meaning $\epsilon_{\textrm{osc}}=1/3$ for our calculation \cite{julia1}.
\newline
The function $F_{\nu}(x,E_{\nu}/x)$ gives the muon neutrino spectrum for a fixed proton energy  $E_{\textrm{p}}$ and is divided into three summands, $F_{\textrm{1}}$, $F_{\textrm{2}}$ and $F_{\textrm{3}}$. The first one denotes  the muon neutrino spectrum of neutrinos from direct pion decay, $F_{\textrm{2}}$ gives the spectrum of neutrinos produced by muon decay and $F_{\textrm{3}}$ considers the electron neutrinos produced \cite{kelner}.
\newline
Further, $\textrm{d}N_{\textrm{p}}/(\textrm{d}A\, \textrm{d}E_{\textrm{p}})$, the incident proton spectrum in units of $\textrm{cm}^{-2}\,\textrm{TeV}^{-1}$, is given by
\begin{align}
\frac{\textrm{d} N_{\textrm{p}}}{\textrm{d}A\,\textrm{d}E_{\textrm{p}}}=\frac{A_{\textrm{p}}}{E_{\textrm{p}}^{p}}\exp\left(-\frac{E_{\textrm{p}}}{E_{0}}\right).
\end{align}
Here, $A_{\textrm{p}}$ is the normalization of the spectrum in units of  $\textrm{cm}^{-2}\,\textrm{TeV}^{-1}$, $E_{0}$ is the cut-off energy and $p$ is the spectral index. In the following calculations, we use $p=2$. While the cosmic ray spectrum might indeed deviate from an $E^{-2}$ behavior (see e.g. \cite{meli}, IceCube point source limits are usually given for an $E^{-2}-$spectrum only \cite{abbasi}. In future work, it will be interesting to investigate the effect of different spectral indices as well, when IceCube limits are provided for other cases as well. For the two sources Cen A and M 87, where $\gamma$-ray measurements indicate  a deviation from an $E^{-2}-$behavior, we actually do use the observed values, as we do not rely on IceCube limits in that case. The cosmic ray normalization is connected to the total cosmic ray energy $W_{\textrm{p}}$ via
\begin{align}
&\int_{E_{p}^{\textrm{min}}}^{E_{p}^{\textrm{max}}}{\frac{\textrm{d} N_{\textrm{p}}}{\textrm{d}A\,\textrm{d}E_{\textrm{p}}}E_{\textrm{p}}\,}\textrm{d}E_{\textrm{p}}=\frac{W_{\textrm{p}}}{4\pi d_{L}(z)^{2}}\\
&\int_{E_{\textrm{p}}}{A_{\textrm{p}}\left(\frac{\textrm{TeV}}{E_{\textrm{p}}}\right)^{p}}\exp\left(-\frac{E_{\textrm{p}}}{E_{0}}\right)E_{\textrm{p}}\,\textrm{d}E_{\textrm{p}}=\frac{W_{\textrm{p}}}{4\pi d_{L}(z)^{2}},\\
&A_{\textrm{p}}=\frac{W_{\textrm{p}}}{4\pi d_{L}(z)^{2}}\cdot\underbrace{\left(\int_{E_{\textrm{p}}}{\left(\frac{\textrm{TeV}}{E_{\textrm{p}}}\right)^{p}}\exp\left(-\frac{E_{\textrm{p}}}{E_{0}}\right)E_{\textrm{p}}\,\textrm{d}E_{\textrm{p}}\right)^{-1}}_{\beta}\\
&A_{\textrm{p}}=\frac{W_{\textrm{p}}}{4\pi d_{L}(z)^{2}}\,\beta.
\end{align}
The cosmic ray emission is  isotropic, meaning that only the fraction $(4\pi d_{L}(z)^{2})^{-1}$, with $d_{L}$ representing the redshift dependent luminosity distance, reaches the Earth.
Due to the scaling law of the cross section with the nuclei number, the results apply for different cosmic ray compositions. We perform the calculations for protons only for simplicity, but expect the same results for a heavier composition.
The minimum energy of the cosmic rays is the threshold energy for the $\pi^{+}$ production, $E_{\textrm{p}}^{\textrm{min}}\approx 1.2\cdot10^{-3}\,\textrm{TeV}$, and $E_{\textrm{p}}^{\textrm{max}}$ is the maximum cosmic ray energy. $E_{\textrm{p}}^{\textrm{min}}\approx 10^{9}$ TeV.
\newline
The total proton energy of a single source is given by
\begin{align}
W_{\textrm{p}}&=\int{L_{\textrm{p}}}\,\textrm{d}t=L_{\textrm{p}}\int{\textrm{d}t}\label{11}\\
&=t_{\textrm{H}}L_{\textrm{radio}}\frac{1}{f_{\textrm{e}}}\int_{0}^{z}{\frac{\textrm{d}z^{\prime}}{(1+z^{\prime})E(z^{\prime})}}.\label{22}
\end{align}
In Equation (\ref{11}) we assume a constant proton luminosity $L_{\textrm{p}}$. The parameter $t_{\textrm{H}}$ is the Hubble time and $E(z)$ is given by \cite{hogg}
\begin{align}
E(z)=\sqrt{\Omega_{\textrm{M}}(1+z)^{3}+\Omega_{\textrm{k}}(1+z)^{2}+\Omega_{\Lambda}}.
\end{align}
In Equation (\ref{22}) we assume a ratio $f_{\textrm{e}}$ between radio and proton luminosity, given by
\begin{align}
f_{\textrm{e}}=\frac{L_{\textrm{radio}}}{L_{\textrm{p}}}=\frac{\int_{E_{\textrm{e}}}{\frac{\textrm{d}N_{\textrm{e}}}{\textrm{d}E_{\textrm{e}}}E_{\textrm{e}}}\,\textrm{d}E_{\textrm{e}}}{\int_{E_{\textrm{p}}}{\frac{\textrm{d}N_{\textrm{p}}}{\textrm{d}E_{\textrm{p}}}E_{\textrm{p}}}\,\textrm{d}E_{\textrm{p}}}.
\end{align}
Considering the mentioned assumptions the normalization is
\begin{align}
A_{\textrm{p}}^{\textrm{Earth}}&=\frac{W_{\textrm{radio}}}{4\pi d_{L}(z)^{2}}\beta\frac{1}{f_{\textrm{e}}}.
\end{align}
\section{Limits on the target densities for 33 FR-I galaxies}\label{FR1}
Fanaroff and Riley divided  radio galaxies according to the correlation between morphology and luminosity into two groups, FR-I and FR-II \cite{fanaroff}. FR-I are brightest towards the center, while FR-II are brightest at outermost part of the jet. Later it was discovered that this behavior is correlated with a critical radio luminosity at 178 MHz, $L_{178}=2\cdot10^{26} \ \textrm{W}/\textrm{Hz}$, dividing the radio sources in FR-I and FR-II, where FR-I have lower and FR-II higher luminosities \cite{chiaberge}. Detailed observation showed that the morphology is correlated to the energy transport in the sources. FR-I sources have bright knots along the jets, while FR-II have faint jets but bright hot spots at the end of the lobes, indicating that they appear to be able to transport energy efficiently to the ends of the lobes. FR-I sources on the other hand are inefficient in the way that a large amount of energy is radiated. Considering the AGN unification scheme, FR-I galaxies are assumed to be the misaligned counterparts of BL Lacs \cite{megan}, meaning that the non thermal beamed emission from the relativistic jets should be present in radio galaxies. Observation show that the ratio of nuclear luminosities of FR-I and BL-Lacs show a correlation with the orientation of FR-I galaxies, supporting the assumption that they are correlated with each other.
\newline
In this paper we focus on FR-I galaxies  presented \cite{chiaberge}. The selection includes 33 FR-I galaxies with a given radio luminosity $L_{\textrm{radio}}$ at 178 MHz. The redshift ranges between $z=0.0037-0.29$. Using the radio luminosity, the redshift $z$, considering the cosmology $(H_{0}=75\ \textrm{km}\ \textrm{s}^{-1}\,\textrm{Mpc}^{-1} \ \Omega_{\textrm{M}}=0.27, \ \Omega_{\Lambda}=0.73)$ and $f_{\textrm{e}}$ the normalization $A_{\textrm{p}}$ and the muon neutrino flux are calculated. The latter can  then be used to calculate the neutrino flux in dependence of the target density $n_{\textrm{H}}$.
\subsection{Results}
Table \ref{table1} provides the upper limits for the target densities of the 33 FR-I galaxies presented in \cite{chiaberge}. Here, the radio luminosity of each single source, the cosmology dependent parameters, luminosity distance $d_{\textrm{L}}(z)$ the lookback time, and the constant ratio $f_{\textrm{e}}=0.1$, were used to compute the normalization $A_{\textrm{p}}$. The resulting muon neutrino fluxes are calculated and matched in normalization to the IceCube limit, resulting in an upper limit to the target density.
\newline
Including the knowledge on neutrino flux limits provided by IceCube \cite{abbasi} allows thus to set  an upper limit on the target density. As IceCube limits are only provided directly for a fixed number of sources, we use the declination dependent sensitivity from \cite{abbasi}.
Since we have detailed information about the high energy part of CRs, we can use observations of data to calculate the proton target density. The  two assumptions we use is that a) UHECRs originate from AGN and b) protons are accelerated in the jet
in the same way as electrons. Taking this assumption we calculate $n_{\textrm{H}}$, which ranges between 20 $\textrm{cm}^{-3}$ for 3C 028 and $1500\,\textrm{cm}^{-3}$ for 3C~386 for $f_{\textrm{e}}=0.1$.
\newline
Comparing our results with  the model of Kazanaz and Elliason \cite{kazanaz}, where a spherically symmetric accretion shock, accelerating a fraction of the inflowing plasma to the highest energies is considered, the proton density is given by
\begin{align}\label{density}
n_{\textrm{H}}(x)\approx 1.15\cdot 10^{9} \frac{\dot{m}}{M_{9}^{2}}x^{-3/2}\, \textrm{cm}^{-3}.
\end{align}
Here, $\dot{m}:=\dot{M}/(1\,M_{\odot}\,\textrm{yr}^{-1})$ is the accretion rate, $M_{9}:=M/(10^{9}\,M_{\odot})$ the black hole mass in units of $10^{9}$ solar masses. The parameter $x:=r/r_{\textrm{S}}$, with  the Schwarzschild radius $r_{\textrm{S}}$, gives the radial distance. Comparing Equation (\ref{density}) with our results indicates that accelerated protons  might originate from a maximum orbit $x_{\textrm{max}}\approx 3\cdot10^{4}$ for $\dot{m}=0.1$, from $x_{\textrm{max}}\approx 5\cdot10^{4}$ for $\dot{m}=0.2$ and from $x_{\textrm{max}}\approx 9\cdot10^{4}$ for $\dot{m}=0.5$ and from a minimum orbit  $x_{\textrm{min}}\approx 2000$ for $\dot{m}=0.1$, from $x_{\textrm{min}}\approx 3000$ for $\dot{m}=0.2$ and from $x_{\textrm{min}}\approx 6000$ for $\dot{m}=0.5$.
\begin{figure}[htbp]
	\centering
		\includegraphics[scale=0.37]{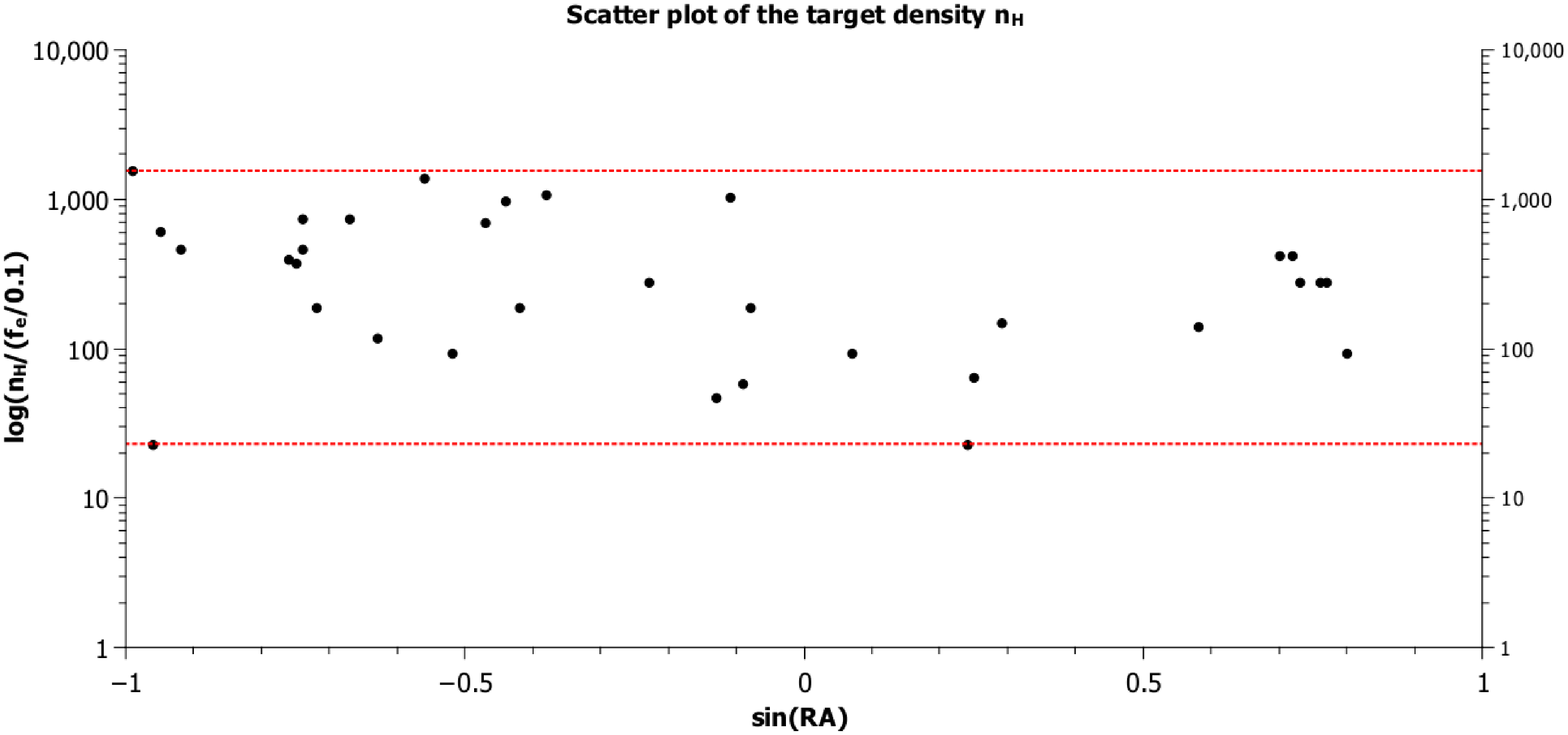}
		\includegraphics[scale=0.35]{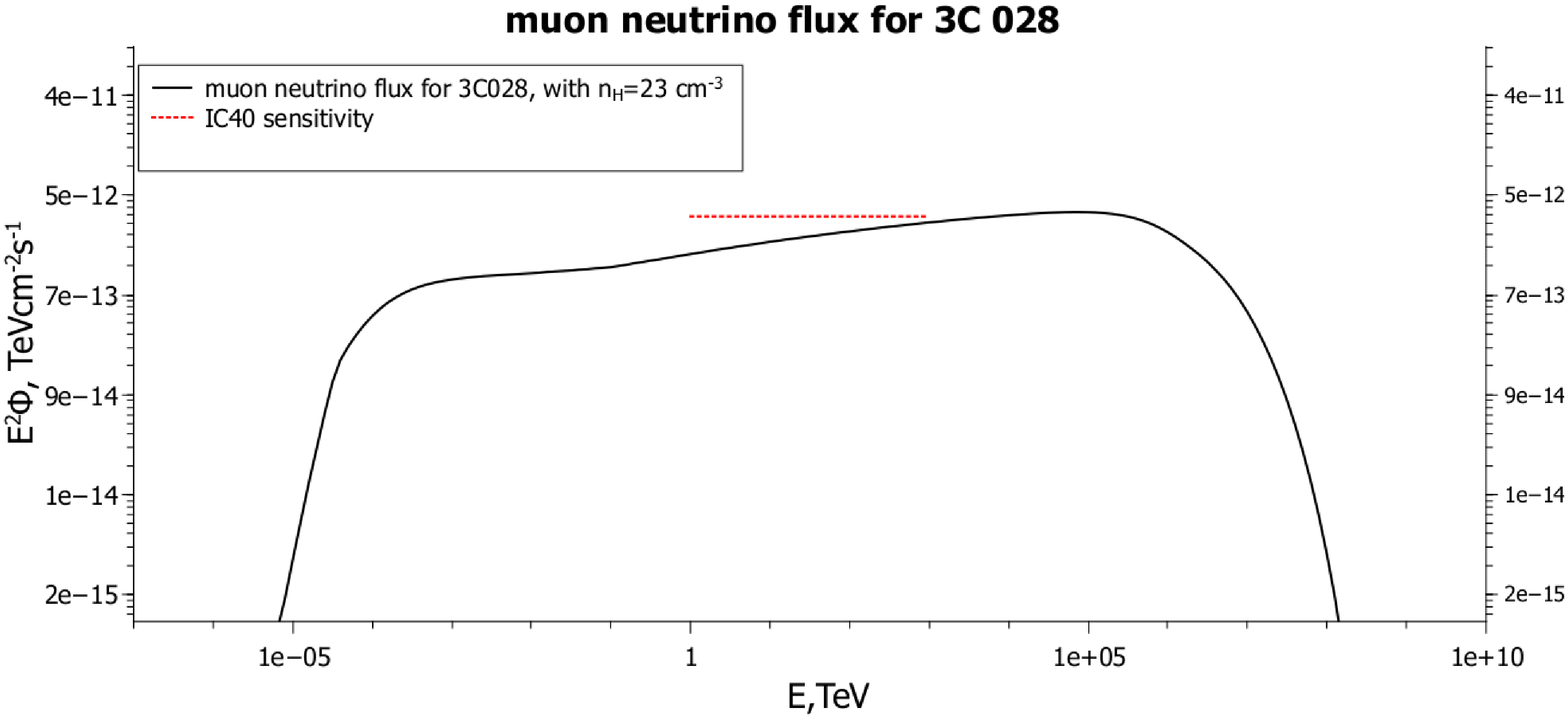}
	\caption{\small scatter plot of  $n_{\textrm{H}}/(f_{\textrm{e}}/0.1)$ as a function of sin(RA) (top) and the muon neutrino flux for 3C 028 (bottom). The red (dashed) lines in the scatter plot give the density range. The red (dotted) line gives the sensitivity of IC40. We show the muon neutrino flux for one source. The other sources have the same shape for the flux but a different value for the sensitivity.}
	\label{fig:proton_density}
\end{figure}
\newline
Considering the jet disk model \cite{falcke1,falcke}, the target density in the jet in the observers frame, is given by
\begin{align}
n_{\textrm{H}}=11\cdot\Gamma\,  L_{46}\,q_{j/1}Z_{\textrm{pc}}^{-2}\, \textrm{cm}^{-3}.
\end{align}
Here, $L_{46}$ is the disk luminosity in units of $10^{46}$ erg/s, $q_{j/1}$ is the ratio between jet power and disk luminosity, $Z_{\textrm{pc}}$ is the distance from the origin in parsec and $\Gamma$ is the boost factor of the plasma. The parameters for this model are $\Gamma\approx10$, $L_{46}\leq 10^{-3}$ and $q_{j/1}\approx0.15$  for  $Z_{\textrm{pc}}< 1$. The resulting densities of $n_{\textrm{H}}\leq1700\,(Z_{\textrm{pc}}/0.1)^{-2}\,\textrm{cm}^{-3}$ are well consistent with the limits derived here and would only start to become inconsistent for $Z_{\textrm{pc}}\ll 0.1$, a distance smaller than $10^{2}\,r_{\textrm{s}}$. So far, it is not clear, how well the jet disk scenario works for FR-I galaxies, though, as their accretion disks are extremely faint and difficult to observe \cite{falcke2}.
\begin{table}[htbp]
\small
\begin{tabular}{|c|c|p{2.7cm}|p{2.7cm}|p{2.2cm}|}\hline
\centering
Name &   sin(RA) &sensitivity$\times 10^{-12}$ \newline in $\textrm{TeV}\textrm{cm}^{-2}\textrm{s}^{-1}$& Normalization in \newline $\textrm{TeV}^{-1}\,\textrm{cm}^{-2}$& $n_{\textrm{H}}/(f_{\textrm{e}}/0.1)$\newline in $\textrm{cm}^{-3}$ \\ \hline\hline
3C 028	&	0.24	&	3.34	&	287.16	&	$<$23 \\ \hline
3C 029	&	0.25	&	3.34	&	101.18	&	$<$65\\ \hline
3C 031	&	0.29	&	3.54	&	48.93	&	$<$149\\ \hline
3C 066B	&	0.58	&	5.27	&	88.80	&	$<$140\\ \hline
3C 075	&	0.70	&	6.42	&	40.52	&	$<$420\\ \hline
3C 76.1	&	0.72	&	6.82	&	63.21	&	$<$420\\ \hline
3C 078	&	0.73	&	7.09	&	91.76	&	$<$280\\ \hline
3C 083.1	&	0.76	&	7.68	&	109.26	&	$<$280\\ \hline
3C 084	&	0.77	&	7.53	&	112.96	&	$<$280\\ \hline
3C 089	&	0.80	&	8.15	&	296.93	&	$<$93\\ \hline
3C 264	&	0.07	&	3.28	&	77.80	&	$<$93\\ \hline
3C 270	&	-0.08	&	7.15	&	58.71	&	$<$187\\ \hline
3C 272.1	&	-0.11	&	9.46	&	9.76	&	$<$1027\\ \hline
3C 274	&	-0.13	&	13.03	&	457.65	&	$<$47\\ \hline
3C 277.3	&	-0.23	&	26.23	&	98.58	&	$<$280\\ \hline
3C 288	&	-0.42	&	52.85	&	348.95	&	$<$187\\ \hline
3C 293	&	-0.47	&	68.55	&	84.06	&	$<$700\\ \hline
3C 296	&	-0.56	&	90.74	&	45.78	&	$<$1400\\ \hline
3C 305	&	-0.67	&	117.76	&	95.14	&	$<$747\\ \hline
3C 310	&	-0.72	&	130.19	&	418.69	&	$<$187\\ \hline
3C 314.1	&	-0.74	&	132.83	&	144.95	&	$<$747\\ \hline
3C 315	&	-0.75	&	143.88	&	224.66	&$<$373	\\ \hline
3C 317	&	-0.76	&	141.05	&	241.42	&	$<$397\\ \hline
3C 338	&	-0.92	&	206.47	&	216.36	&	$<$467\\ \hline
3C 346	&	-0.95	&	223.69	&	175.10	&	$<$607\\ \hline
3C 348	&	-0.96	&	215.00	&	5529.19	&	$<$23\\ \hline
3C 386	&	-0.99	&	257.34	&	68.99	&	$<$1552\\ \hline
3C 424	&	-0.74	&	143.88	&	200.83	&	$<$467\\ \hline
3C 433	&	-0.63	&	100.33	&	699.21	&	$<$117\\ \hline
3C 438	&	-0.52	&	72.81	&	896.32	&	$<$93\\ \hline
3C 442	&	-0.44	&	59.59	&	70.93	&	$<$968\\ \hline
3C 449	&	-0.38	&	43.26	&	38.05	&	$<$1073\\ \hline
3C 465	&	-0.09	&	7.60	&	185.54	&	$<$58\\ \hline
\end{tabular}
\caption{The observed 33 FR-I galaxies. Column 1 gives the the names, column 2 the sine of right  ascension, column 3 gives the sensitivity of IC40, column 4 the normalization and column 5 the target density for $f_{\textrm{e}}=0.1$}
\label{table1}
\end{table}
\section{Cen A and M 87}\label{fermI}
\begin{figure}[htbp]
	\centering
		\includegraphics[width=0.90\textwidth]{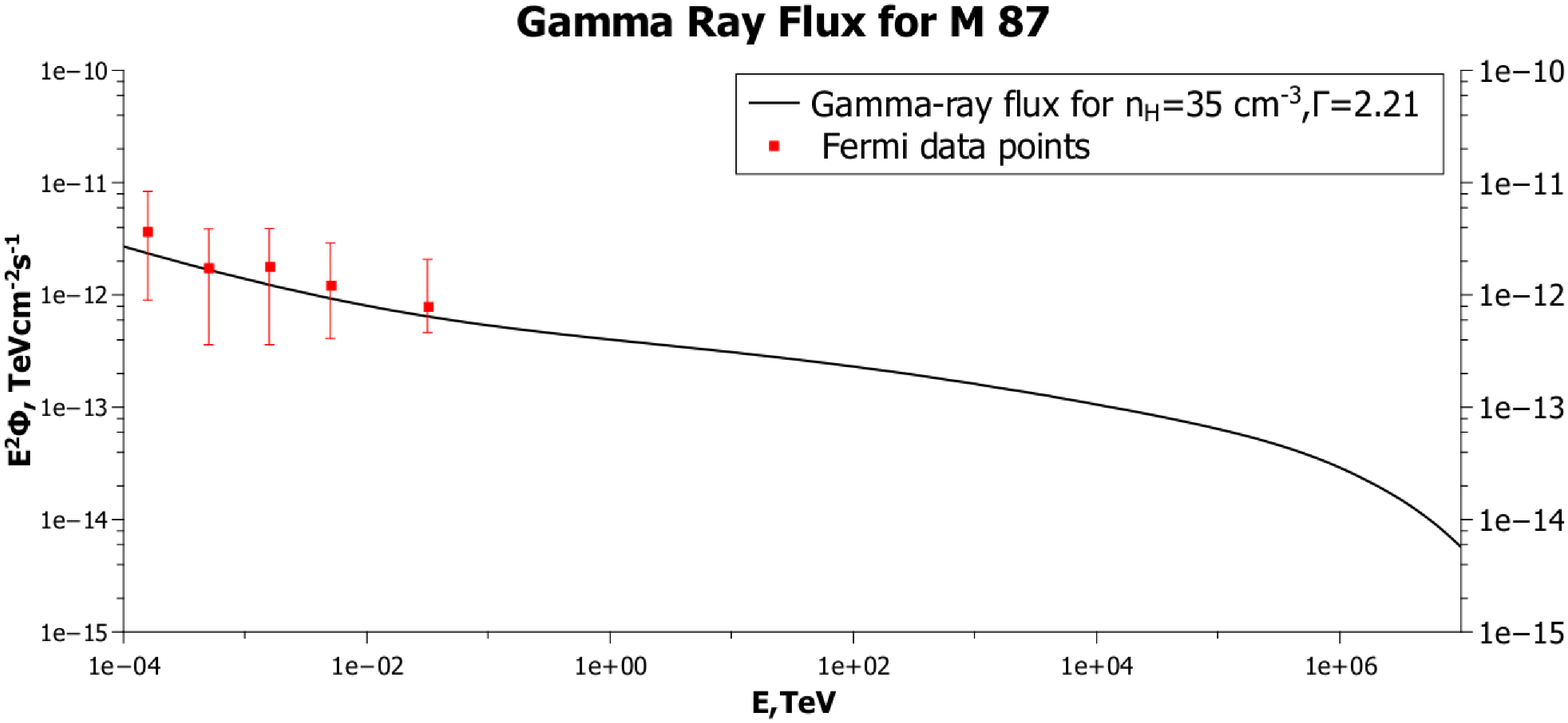}
	\includegraphics[width=0.90\textwidth]{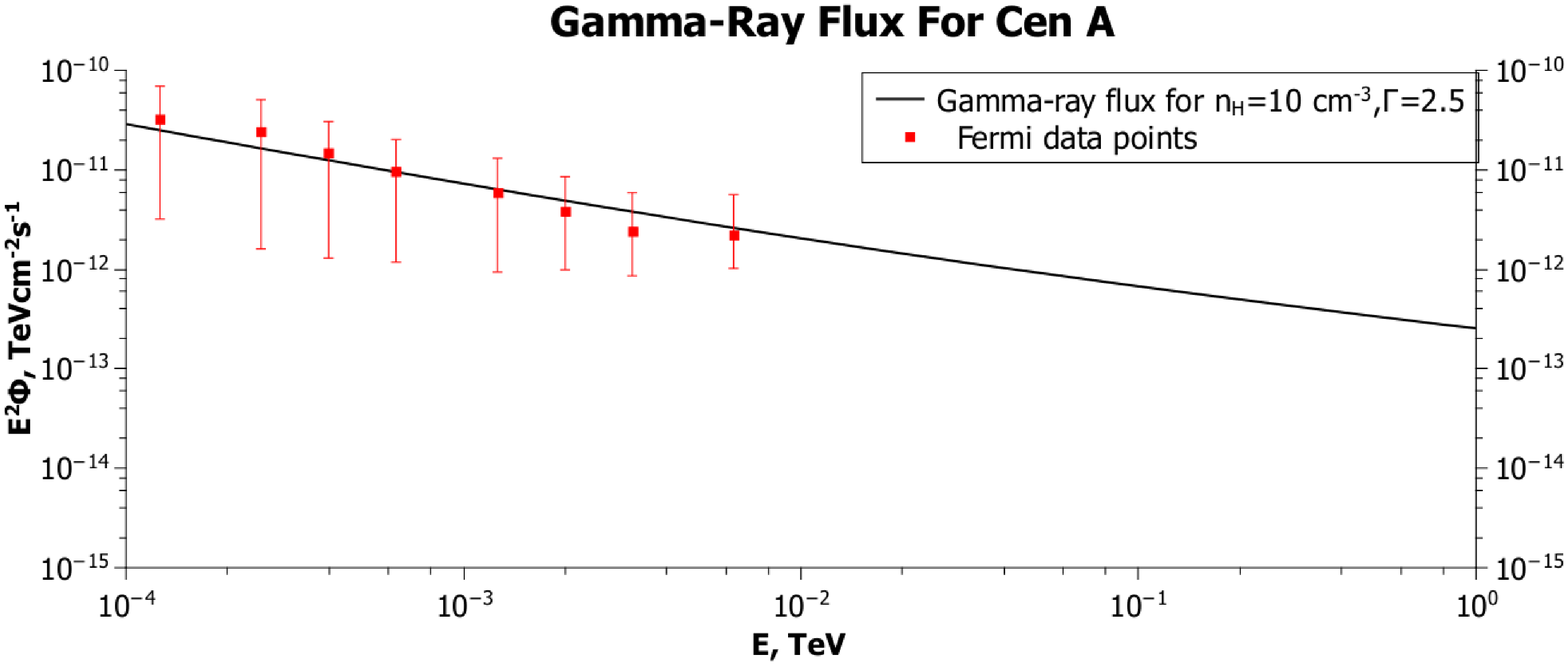}
	\caption{Gamma ray flux for M 87 (top) and for Cen A (bottom). The (red) points give the Fermi observation \cite{abdo}. The spectral index for the proton spectrum is $\Gamma$.}
	\label{fig:m87gamma_mitfehler}
\end{figure}

For Cen A and M 87 we use the Fermi \textsl{LAT} observations to normalize the $\gamma$-ray flux, which fixes the neutrino flux directly. Thus IceCube limits are not needed in this case. The Large Area telescope is a pair conversion $\gamma$-ray telescope, covering the energy range from 20 MeV to more than 300 GeV \cite{abdo}. Due to its vicinity to Earth, Cen A has been well studied over the entire electromagnetic spectrum, from radio to $\gamma$-rays. Observations performed by experiments like the Auger observatory indicate that the origin of the highest energy CRs ($E\geq10^{19}$eV) could correlate with the angular position of Cen A \cite{abraham}.
\newline
M 87  one of the nearest $(d=16\  \textrm{Mpc})$ and best studied radio galaxies, just like Cen A is known for its bright arcsec-scale jet. It contains an SMBH with a mass~of~$\sim(3-6)\cdot10^{9}\,\textrm{M}_{\odot}$. At TeV energies M 87 is detected by H.E.S.S. \cite{aha}, MAGIC\cite{alb} and VERITAS \cite{acc}.
\newline
Due to the short observation period of MAGIC, VERITAS and H.E.S.S. in comparison to Fermi \textit{LAT} we focus on Fermi observations for M87. A further reason is the time variability of the $\gamma$-flux for $E> 730$ GeV \cite{aha}, while Fermi gives due to the longer observation period  time averaged fluxes. A  satisfactory answer to the question  of the source of high energy $\gamma$-rays is not yet found. %Taking the 95\% confidence error radius $r_{95\%}=0^{\circ}.086$ \cite{abdo2} the diameter can be estimated.
\newline
For Cen A we also focus on Fermi observations, since Fermi \textit{LAT} can resolve the inner region of Cen A \cite{abdo1}.
A further reason is  the $\gamma\gamma$  absorption, making it unlikely that the $\gamma$-ray flux measured by H.E.S.S. and Fermi, originates from the same region \cite{cena}.
% After Cen A and NGC1275, M 87 is the third radio galaxy observed with Fermi \cite{abdo}.
\newline
First we give an estimate on the target density for the two sources, by assuming that $\gamma$-rays are produced in the inner region, leading to the condition
\begin{align}
\lambda=\frac{1}{\sigma_{\textrm{pp}}n_{\textrm{H}}}=\textrm{D}_{\textrm{source}}.
\end{align}
Here, $\lambda$ is the mean free path and $\textrm{D}_{\textrm{source}}$ is the diameter of the source. Using this assumption excludes the other AGN from \cite{abdo}, since the origin of the observed $\gamma$-rays cannot be absolutely restored to the inner region of the sources.
\newline
For the observed energy range, the cross section remains almost constant see (Equation (\ref{sigma}))
\begin{align}
\lambda&=\frac{1}{\sigma n_{\textrm{H}}}\\
&\approx \frac{1}{30 \ \textrm{mb}}\frac{1}{n_{\textrm{H}}}\\
&\approx 9.67 \ \textrm{pc} \left(\frac{10^{6} \ \textrm{cm}^{3}}{n_{\textrm{H}}}\right).
\end{align}
Assuming that the cores are spherical, using the distance to the source, the diameter of the cores can be estimated, as
\begin{align}
\textrm{D}_{\textrm{ Cen A}}&=1.5\cdot 10^{5} \ \textrm{pc}\\
\textrm{D}_{\textrm{ M 87}}&=2.3\cdot10^{4} \ \textrm{pc}.
\end{align}
For M 87 the 95\% confidence error radius $r_{95\%}=0^{\circ}.086$ is used \cite{abdo2}.
%For the calculation of the densities the cross section $\sigma_{\textrm{inel.}}$ is required. For the observed energy range it remains almost constant [Becker Biemann 2009] and can be used to approximate the  mean free path $\lambda$
Considering our assumptions, the mean free path has to fulfill the following relation,
\begin{align}
\lambda&=\textrm{D}_{\textrm{source}}\label{cs}\\
&\Rightarrow  n_{\textrm{H}}\left(f_{\textrm{e}}/0.1\right)^{-1}=70 \, \textrm{cm}^{-3} \, \textrm{for Cen A}\label{cs1}\\
&\Rightarrow  n_{\textrm{H}}\left(f_{\textrm{e}}/0.1\right)^{-1}=430 \, \textrm{cm}^{-3} \, \textrm{for M 87}.
\end{align}
It has to be noticed that the assumptions used here are  simplifications. For example, the geometry can be very complex due to physical processes, like clumping, so that the assumption of  spherical geometry might not be sufficient to describe the real shape of the source. This has an influence on the target density, Equation (\ref{cs}). Furthermore we assume $\gamma$-production only in the inner region. Nevertheless the above calculation provides a first order approximation of the required target densities.
\newline
We compare the results from the estimate with the results, using our model explained in section 2. Therefore, we use the radio luminosity of the inner region to calculate the normalization. The fraction $1/10=L_{\textrm{core}}^{\textrm{radio}}/L_{\textrm{total}}^{\textrm{radio}}$, observed for Cen A given in \cite{alvarez} is assumed to be the same for M 87. Using a power law function, with $\Gamma$ for the spectral index and $E_{0}$ for the  cut-off energy, a power-law spectrum of primary cosmic rays is used to model the $\gamma-$spectrum:
\begin{align}
\frac{\textrm{d}N}{\textrm{d}E\,\textrm{d}A}&=A_{\textrm{p}}E^{-\Gamma}\exp(-E/E_{0})
\end{align}
Here, we used $E_{0}= 10^{8}$TeV as the maximum energy, which is needed to explain the observed flux of ultra-high energy cosmic rays. The Fermi data for Cen A and M 87 require spectral indices of $\Gamma_{\rm Cen A} = 2.5$ and $\Gamma_{\rm M 87} = 2.21$.
Considering the cosmology, the radio core luminosity and the distances, we calculate the normalization $A_{\textrm{p}}$, the $\gamma$-ray flux (Figure \ref{fig:m87gamma_mitfehler}) and the densities for  $f_{\textrm{e}}$=0.1:
\begin{align}
A_{\textrm{Cen A}}^{\textrm{p}}&= 459\, \textrm{TeV}^{-1}\,\textrm{cm}^{-2}\, &\Rightarrow  n_{\textrm{H}}\left(f_{\textrm{e}}/0.1\right)^{-1}&=10 \ \textrm{cm}^{-3}\label{cs2}\\
A_{\textrm{M 87}}^{\textrm{p}}&= 136\, \textrm{TeV}^{-1}\,\textrm{cm}^{-2}\, &\Rightarrow  n_{\textrm{H}}\left(f_{\textrm{e}}/0.1\right)^{-1}&=35 \ \textrm{cm}^{-3}.
\end{align}
As can be seen for Cen A the density estimated from the core size, Equation (\ref{cs1}) and the one calculated within the model presented, Equation (\ref{cs2}) are in agreement.
\newline
For M 87 the situation is different. Uncertainties like the unknown core size or the unknown core luminosity have a large influence on the estimate. To eliminate these uncertainties, the core size has to be resolved and the fraction of the core to the whole radio luminosity has to be known. For this purpose, improved measurements for the core size of M 87 will help. Additionally, the detailed observations of a larger sample of AGN in the future can contribute to receive a statistical sample of the fraction of core to total luminosity of FR-I sources. If the variance is of physical nature, on the other hand, the ratio $f_{\textrm{e}}$ might deviate significantly from $f_{\textrm{e}}=0.1$.
\subsubsection{The density and muon neutrino spectrum for Cen A and M 87}
The muon neutrino flux can be derived directly from the fit to the $\gamma$-measurement and compared with the limits given in \cite{abbasi}.
As one can see, the flux is below the current sensitivity of the IC40 for both sources. It has to be mentioned that the uncertainties in the assumptions are significant, so the results have to be interpreted carefully.
For M 87 the sensitivity is about one order of magnitude and for Cen A two orders of magnitude higher than the flux in the corresponding energy range. Long-term observations with IceCube and possible future extensions might be able to yield a first significant signal from M 87. In the case of Cen A, the measured flux is simply too steep to be obesrvable by high-energy neutrino telescopes. Other emission regions, like outer parts of Cen A, might be more interesting and existing H.E.S.S.\ measurements reveal the possibility of such sources with a rather flat spectrum \cite{hess1}. Future and more sophisticated experiments like KM3NET will help to obtain better observations, since it is optimized for the southern hemisphere and therefor sensitiv to  lower TeV energies \cite{k3net}.
\newline 
 Since the neutrino flux is derived from the $\gamma$-measurements, uncertainties in $\gamma$-observations have an influence on the estimates of the neutrino flux and hence an influence on the difference between calculated neutrino flux and the sensitivity of the detector. The more precise the $\gamma$-observations are, the more precise are the calculated neutrino fluxes, resulting in an improved comparison between sensitivity and the flux. For this reason experiments with improved equipment like more sophisticated satellites and the future ground based Cherenkov detector array (CTA) will help to obtain more detailed $\gamma$-ray observations. If the spectrum of Cen A turns out to be significantly flatter than it is currently indicated by Fermi measurements, the central source might still be interesting for neutrino telescopes. 
\begin{figure}[htbp]
	\centering
		\includegraphics[width=0.90\textwidth]{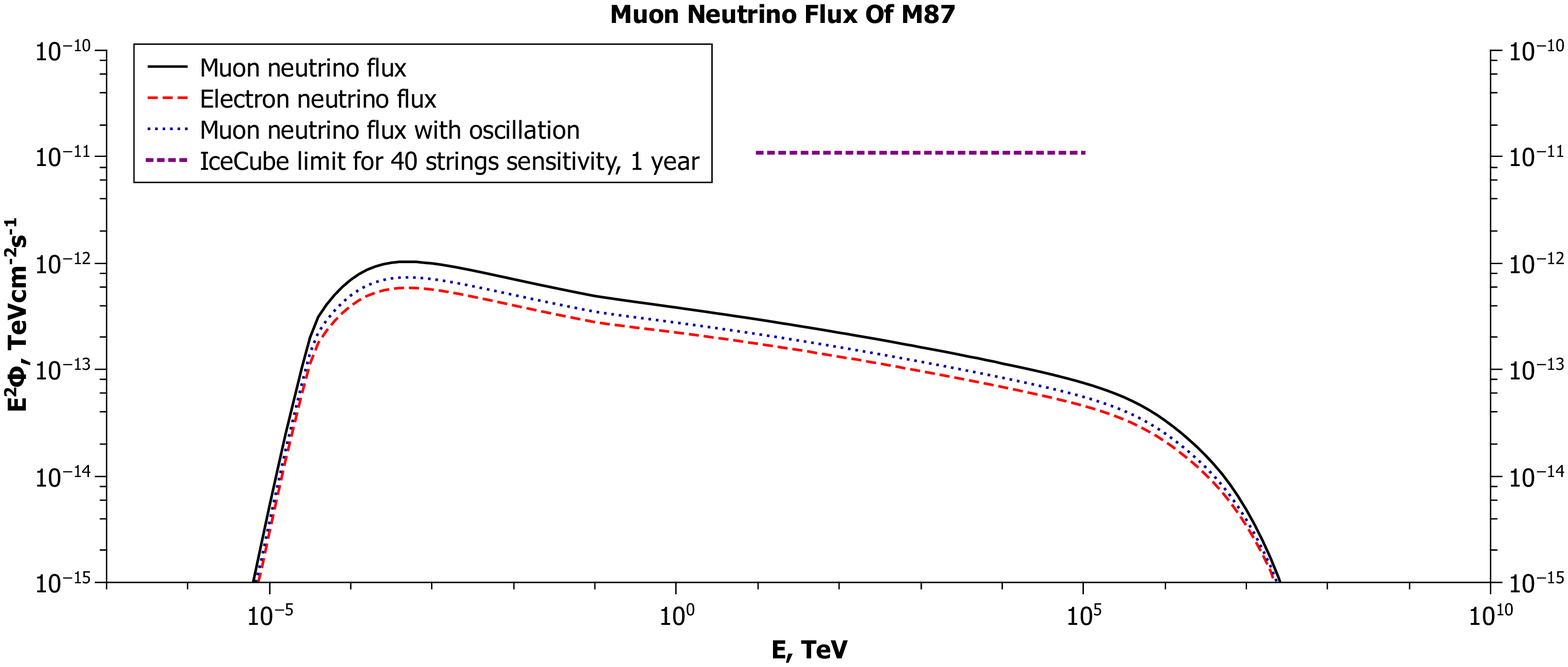}
		\includegraphics[width=0.90\textwidth]{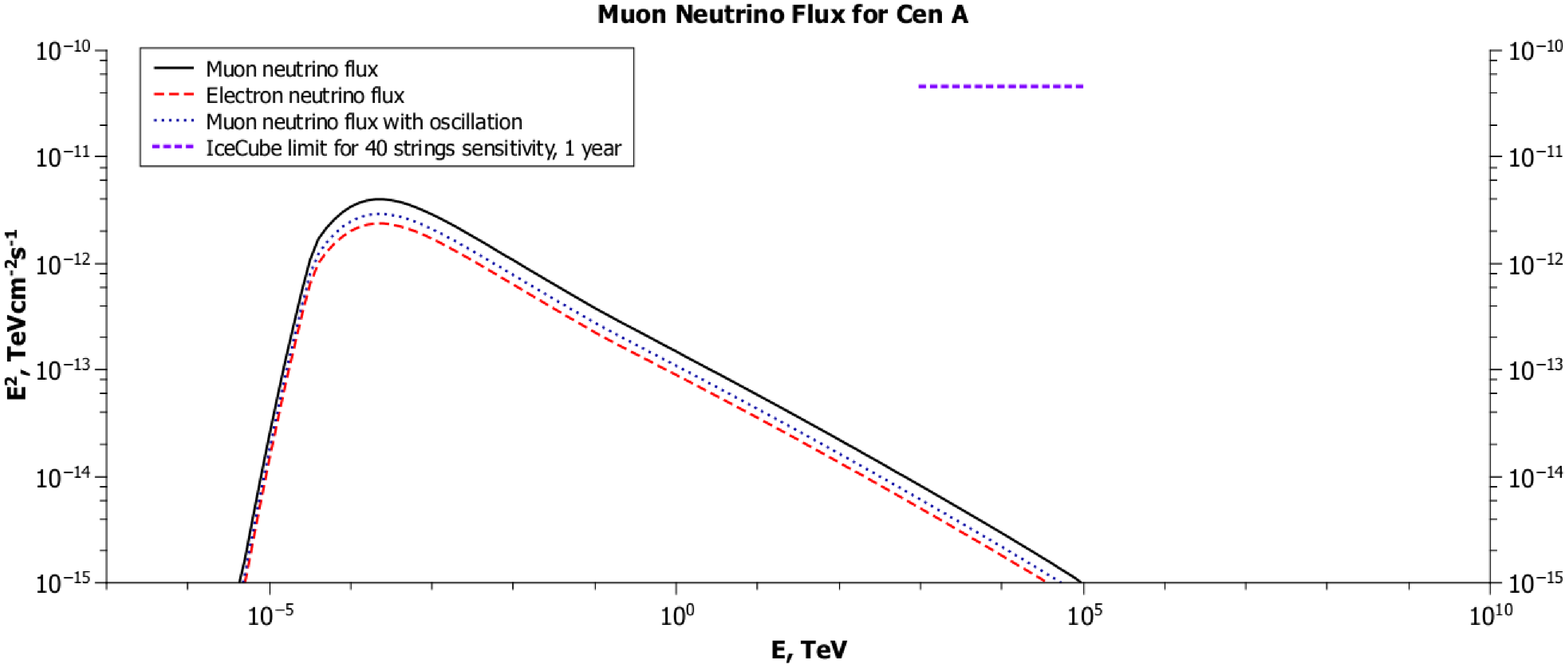}
	\caption{Muon neutrino flux for M 87 and Cen A. The black (solid) curve is the spectrum and the red (dashed) the electron neutrino spectrum. The dark blue (dotted) curve is the muon neutrino spectrum, if oscillations are considered \cite{abdo}. The blue (dashed) horizontal line gives the IceCube sensitivity for  M87 at the northern sky and for Cen A at the southern sky \cite{abbasi}. }
	\label{fig:m87_neutrino}
\end{figure}
\newpage
\section{Conclusions}\label{conclusion}
In this paper, we derive limits on the product $n_{\textrm{H}}/f_{\textrm{e}}/(0.1)$ for 33 FR-I galaxies, using the radio luminosity at 178 MHz and neutrino flux limits. 
The target density can be limited to be between 20 $\textrm{cm}^{-3}$  and $1500\,\textrm{cm}^{-3}$,  for a fixed $f_{\textrm{e}}=0.1$ for the sources. An explanation is the normalization of the sources. Two objects at the same location at the sky but different normalization, differ in the target density, since it is $A_{\nu}\propto n_{\textrm{H}}$, resulting in $A_{\nu_{1}}/A_{\nu{2}}=n_{H_{2}}/n_{H_{1}}$ for the two objects. Two objects with a  different location but approximately same normalization, the ratio is $n_{H_{1}}/n_{H_{2}}\propto\alpha_{1}/\alpha_{2}$, with  $\alpha$ for the sensitivity of the detector for the object. Considering different models of the distribution of matter in AGN, we conclude that the innermost core models for proton acceleration can be excluded, as densities there are expected to be of the order of $10^{9}$~cm$^{-3}$. An alternative explanation would be that the ratio of protons to electrons must be significantly lower than expected from standard theory \cite{krause,schlicky}.
\newline
Gamma-ray observations exist for Cen A and M 87, the two closest and best studied AGN and they are used in this paper to estimate the average density of the emission region in case of hadronic interaction processes. For our calculation we used Fermi observations, since the observation period by Fermi reflects the average flux over a long-time measurement, rather than individual short-term measurements by Imaging Air Cherenkov Telescopes like MAGIC, VERITAS and H.E.S.S. Those provide interesting insights on the flaring behavior of the sources, but no reliable measurement of an average flux at this point. We find that the density required for gamma-ray emission in the core of Cen A and M 87 has to be of the order of $10-100$~cm$^{-3}$. The exact value again depends on the ratio of protons to electrons accelerated at the source.

In the first part we assumed that $\gamma$-rays are only produced in the inner region and that the inner region is approximated by a spherical symmetry. We estimated the target density and compared these results with the densities we obtained by using the model explained in section 2.  For Cen A the differences between the estimate and the model are in agreement within the uncertainty of the measurements, while for M 87 the situation is different. The differences can explained by our assumptions. Due to physical process, like clumping, the geometry can be very complex. Additionally $\gamma$-ray production might not only takes place in the inner region of the objects. In the case of M 87 neither radio luminositiy observations of the core region nor observations of the core region exist. This leads to a bigger difference between estimate and model. Future measurements will give more detailed information about the inner region of the sources and thus help with the identification of the sources of ultra-high energy cosmic rays.

\section*{Acknowledgements}
We acknowledge generous constantly support from many scientists. We would like to thank Florian Schuppan, Matthias Mandelartz for their contribution to finish the paper. We also thank Peter Biermann and Anthony Brown for the very helpful discussions we had.
\newline
Furthermore we would like to thank the IceCube collaboration for grantig information about the detector.
\newline
IS and JBT furthermore acknowledge support from the DFG grant BE-3714-1, `cosmic ray tracers from gas-rich active galaxies', as well as from the Research Department of Plasmas with complex Interactions Bochum.

\end{document}